\documentstyle[prl,aps,epsf,twocolumn]{revtex}
\tighten
\begin{document}
\def\pr{Phys.\ Rev.\ B }
\def\prl{Phys.\ Rev.\ Lett.\ } 
\draft
\title{Quantum Hall effect in spin-degenerate Landau levels: 
Spin-orbit\\
enhancement of the conductivity}
\author{D.\ G.\ Polyakov\cite{DGP}}
\address{Institut f\"ur Theoretische Physik, 
Universit\"at zu K\"oln, Z\"ulpicher Str. 77, 50937 K\"oln, Germany}
\author{M.\ E.\ Raikh}
\address{Department of Physics, University of Utah, 
Salt Lake City, Utah 84112
\vspace{18pt}}
\author{\small\parbox{14.1cm}{\small 
The quantum Hall regime in a smooth random potential is considered
when two disorder-broadened Zeeman levels overlap strongly. Spin-orbit
coupling is found to cause a drastic change in the percolation network
which leads to a strong enhancement of the dissipative conductivity at
finite temperature, provided the Fermi level $E_F$ lies between the
energies of two delocalized states $E=\pm\Delta$, $2\Delta$ being the
Zeeman splitting. The conductivity is shown to exhibit a box-like
behavior with changing magnetic field: $\sigma_{xx}$ is $\sim e^2/h$
at $|E_F|<\Delta$ and exponentially small otherwise. Two peaks of
$\sigma_{xx}$ arising as $T\to 0$ are found to be strongly asymmetric.
}}
\address{\vspace{-15pt}}
\maketitle
\narrowtext

The conventional picture of the integer quantum Hall effect (QHE)
implies that there is only one delocalized state in the middle of a
disorder-broadened Landau level. Numerical simulations
\cite{aoki85,chalker88,hanna94} support this concept: the localization
length $\xi$ is shown to diverge as $|E|^{-\gamma}$, $\gamma\simeq
2.3$, when the energy $E$ approaches the Landau level center
($E=0$). This conclusion can also be drawn from low-temperature
measurements of the longitudinal conductivity $\sigma_{xx}$ in GaAs
heterostructures: the width of a peak of $\sigma_{xx}$ shrinks with
lowering $T$ as $T^\kappa$ down to 25 mK
\cite{wei88,wei92,koch91,kochprl91}. The localization-length exponent
$\gamma\simeq 2.3$ was measured directly by analyzing how the peak
width scales with the sample size in small Hall-bar geometries
\cite{kochprl91}.

The above picture applies when the disorder-induced width of the
Landau level $\Gamma$ is smaller than the Zeeman splitting
$2\Delta$. In the opposite case, $\Gamma\gg\Delta$, one should expect
the existence of two delocalized states within a single peak of the
density of states, each corresponding to a different projection of
spin. As a result, two $\sigma_{xx}$-peaks may correspond to one peak
of the density of states. Whether the $\sigma_{xx}$-peak
spin-splitting is observable is determined by both $T$ and the
strength of disorder: the peaks merge with increasing $T$ at a
characteristic temperature which is a growing function of the
parameter $\Delta/\Gamma$.

The splitting has been recently observed \cite{hwang93} by tilting the
sample with respect to the magnetic field (this technique makes it
possible to increase the effective $g$-factor). The analysis of the
temperature behavior of $\sigma_{xx}$ at different values of
$\Delta/\Gamma$ enabled the authors to conclude that the data for
$\sigma_{xx}$ as a function of the Fermi level position cannot be
represented as a superposition of two single peaks not related to each
other. Namely, confirming previous experimental results
\cite{wei90,kochprl91}, spin-unresolved $\sigma_{xx}$-peaks were
claimed to shrink with decreasing $T$ anomalously slow - with the
exponent $\kappa$ approximately half that for a single Zeeman
level. The anomaly in the behavior of spin-degenerate
$\sigma_{xx}$-peaks has been recently reported also with respect to
their microwave-frequency broadening \cite{engel93} and broadening
with current \cite{wei94}. However, it was argued in subsequent
discussions \cite{lee94,wang94,hanna94} that a wider range of
experimental parameters is needed in order to make a conclusive
statement about the value of $\kappa$. What we would like to note in
this connection is that when the sweeping of the Fermi level shows two
close spin-{\it split} $\sigma_{xx}$-peaks, the width of each of them,
though being characterized by the ``normal" value of $\kappa$, is much
larger than that of well-separated Zeeman peaks at the same $T$. In
other words, if the width of the peak $\Delta\nu$, $\nu$ being the
filling factor, is represented in the form $\Delta\nu=(T/T_1)^\kappa$,
the characteristic temperature $T_1$ for close spin-split peaks is
observed to be much smaller than for nonoverlapping Zeeman levels. For
example, the values of $T_1$ extracted from the data presented in
Ref.\ \cite{wei88} are as follows: $T_1\sim 30K$ for $N=1$ spin-split
peaks whereas it is $\sim 600K$ for $N=0\downarrow$ peak. This
observation seems to be compelling experimental evidence that
overlapping of Zeeman levels indeed can strongly impede the
localization of electron states.

Clearly, the anomalous behavior of $\Delta\nu$ for two close Zeeman
levels can be accounted only for the spin-orbit (SO) interaction. The
question is, How can the weak SO-interaction manifest itself strongly
in the conductivity? The purpose of the paper is to give answer to the
question by considering a quasiclassical model of electron motion in a
long-range random potential. We will show that in this case the
SO-interaction can lead to a drastic change in the percolation network
which causes a strong enhancement of the conductivity in the QHE
regime.

To study the role of the SO-interaction comprehensively, understanding
of the nature of the localization in the QHE regime should have been a
starting point. However, by now an analytical theory of the quantum,
disorder-induced, localization of the Landau level states is
missing. There exists a completely classical approach to the
localization \cite{trugman83}. It pertains to the case of a smooth
random potential $V(\bbox{\rho})$ with a correlation radius $d$ much
larger than the magnetic length $\lambda$. Electrons move along the
equipotential lines $V(\bbox{\rho})=E$, so that their trajectories are
closed. The exception is one of the equipotentials $V(\bbox{\rho})=0$
which penetrates through the entire system. In the classical picture,
only electrons on this percolating trajectory contribute to
$\sigma_{xx}$ at $T=0$. As it was first pointed out by Chalker and
Coddington \cite{chalker88}, the picture has the defect that tunneling
through saddle-points of $V(\bbox{\rho})$ is ignored. Meanwhile this
tunneling becomes crucial in an energy band of the width
$\Delta_t\sim\Gamma (\lambda/d)^2$ \cite{tunwidth} around the level
$E=0$ causing the coupling of electron states in adjacent cells of the
percolation network.

Since $\Delta_t\ll\Gamma$ in the smooth potential, a plausible
situation is that the Zeeman splitting \cite{zs}, being much smaller
than $\Gamma$, is still much larger than $\Delta_t$
($\Delta_t\ll\Delta\ll\Gamma$). In other words, the Zeeman levels may
overlap while the tunneling through the saddle-points may be
neglected. It is this case that is considered in the Letter (the
opposite case, $\Delta\alt\Delta_t$, has been considered numerically
in Refs.\ \cite{lee94,wang94}). The absence of the tunneling makes it
reasonable to start with the classical picture of electron states. The
prime role of the SO-interaction in that case is illustrated in Fig.\
1 where a pattern of classical trajectories for electrons with spin-up
and spin-down is depicted. All trajectories shown correspond to the
{\it same} energy. They are solutions of the equations
$V(\bbox{\rho})-\Delta=E$ (spin-up) and $V(\bbox{\rho})+\Delta=E$
(spin-down). The trajectories for spin-down are separated by a
saddle-point which is supposed to be non-transparent (its height is
much larger than $\Delta_t$), so that two spin-down states are
decoupled in the absence of the SO-interaction. The crucial
observation is that if the height of the barrier for spin-down is
smaller than $2\Delta$, then the trajectories with different spins are
of different topology. Far left from the saddle-point the
SO-interaction couples the left spin-down trajectory to the spin-up
one while far right the coupling of the same spin-up trajectory to the
right spin-down trajectory takes place. So the spin-up trajectory
provides an effective coupling between the two spin-down states. When
some two spin-up trajectories are separated by a saddle-point, a
spin-down state plays the role of ``mediator" between them. As a
result, the SO-interaction promotes delocalization of electron states.

To introduce characteristic lengths, consider the energy $E=0$ in the
middle between two percolation thresholds $E=\pm\Delta$ for two
projections of spin. The equipotentials $V(\bbox{\rho})\pm\Delta=0$
are two sets of closed lines nonintersecting with each other. As we
assume that $\Delta\ll\Gamma$, any of the spin-up equipotentials goes
parallel with a neighboring spin-down one (typically the distance
between them is $\delta\sim d\Delta/\Gamma$), so that ``change of
partners'' can occur only near sparse saddle-points which fall in the
narrow space between two trajectories. At the same time, percolating
trajectories corresponding to either projection of spin form critical
clusters with the same characteristic radius $R(\Delta)\sim
d(\Gamma/\Delta)^{4/3}\gg d$ \cite{isichenko92}. Because the two
percolating networks cannot cross each other, they must share critical
saddle-points (which are the nodes of the percolation network where
the critical clusters with the same spin-projection are closest to
each other). Important to us will be the characteristic length of the
trajectories $L(\Delta)$ between two critical saddle-points. In fact,
the trajectories are very tortuous and $L(\Delta)\sim
d(\Gamma/\Delta)^{7/3}$ \cite{isichenko92} much exceeds
$R(\Delta)$. This length is relevant because it is $L(\Delta)$ that
determines the length of close contact of two neighboring trajectories
with opposite spins. After traveling together this distance they hit a
critical saddle-point which separates them \cite{loop}.
\begin{figure}
\epsfxsize=3.5truein
\vspace{-1.9truein}
\epsffile{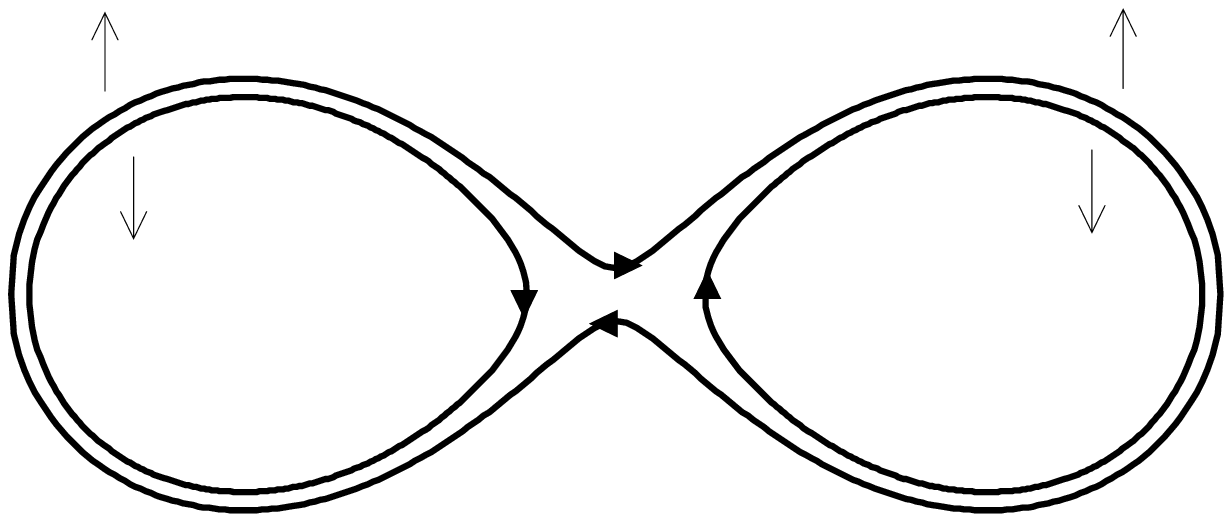}
\vspace{-1.8truein}
\label{fig:loop}
\baselineskip=10pt
{{\small FIG.\ 1.\quad Classical trajectories corresponding to the
same energy, close to that of a saddle-point, in the presence of
spin-splitting. The saddle-point separates two trajectories for
spin-down but does not split up the trajectory for spin-up. Arrows on
the trajectories indicate direction of motion. Due to the spin-orbit
interaction, an effective coupling between the spin-down states is
provided by the spin-up trajectory without any tunneling through the
saddle-point.}}
\end{figure}
We characterize the strength of SO-coupling by the length
$L_{so}$. Its physical meaning is that an electron wave packet which
is initially on, say, a spin-up trajectory will be equally distributed
among the spin-up and spin-down trajectories typically after traveling
the length $L_{so}$. According to the picture above, the crucial
parameter is the ratio $L_{so}/L(\Delta)$. Evaluation of $L_{so}$ in
the case of a smooth random potential will be published elsewhere.

The question we now turn to is, Suppose $L_{so}\alt L(\Delta)$, how
then will a classical electron travel over the system? Let us
demonstrate that if $|E|<\Delta$, the electron, following the
classical trajectories, can percolate. On the contrary, if $E$ is
outside the band $(-\Delta,\Delta)$, its motion is restricted to a
finite area. First of all, note that the trajectories with opposite
spins at a given $E$ may be obtained as the equipotentials
$V(\bbox{\rho})=E\pm\Delta$. Therefore, if one considers a point in
the space between two neighboring trajectories with opposite spins, it
will belong to an equipotential whose energy is somewhere in the
interval $(E-\Delta,E+\Delta)$ and vice versa: all equipotentials with
energies lying in this interval are confined between the spin-up and
spin-down trajectories with the energy $E$. Now let us color in the
area between these trajectories. It is important that at $|E|<\Delta$
the dashed regions form an infinite network [Fig.\ 2(a)]. To prove
this statement, notice that the infinite equipotential
$V(\bbox{\rho})=0$ goes at $|E|<\Delta$ exclusively inside the dashed
space. Therefore, an electron can travel throughout the entire system
following the boundaries of the dashed regions. Treating $R(\Delta)$
as an elementary step we may view the electron motion as a random walk
process. Consider now the case $|E|>\Delta$. Then there is no
percolation at all. Indeed, the dashed area contains now
equipotentials either with only positive or negative energies [Fig.\
2(b)]. In either case, the infinite equipotential corresponding to
zero energy lies outside the dashed space. Hence, passage of an
electron through the sample is inavoidably associated with tunneling
between the finite clusters the characteristic distance between which
is of order $\lambda[(|E|-\Delta)/\Delta_t]^{1/2}$. In other words, if
$|E|$ exceeds $\Delta$ by the small energy $\Delta_t$, the transport
is exponentially suppressed. We thus conclude that the SO-enhancement
of the classical transport occurs in the energy interval $|E|<\Delta$.
\begin{figure}
\epsfxsize=2.8truein
\epsffile{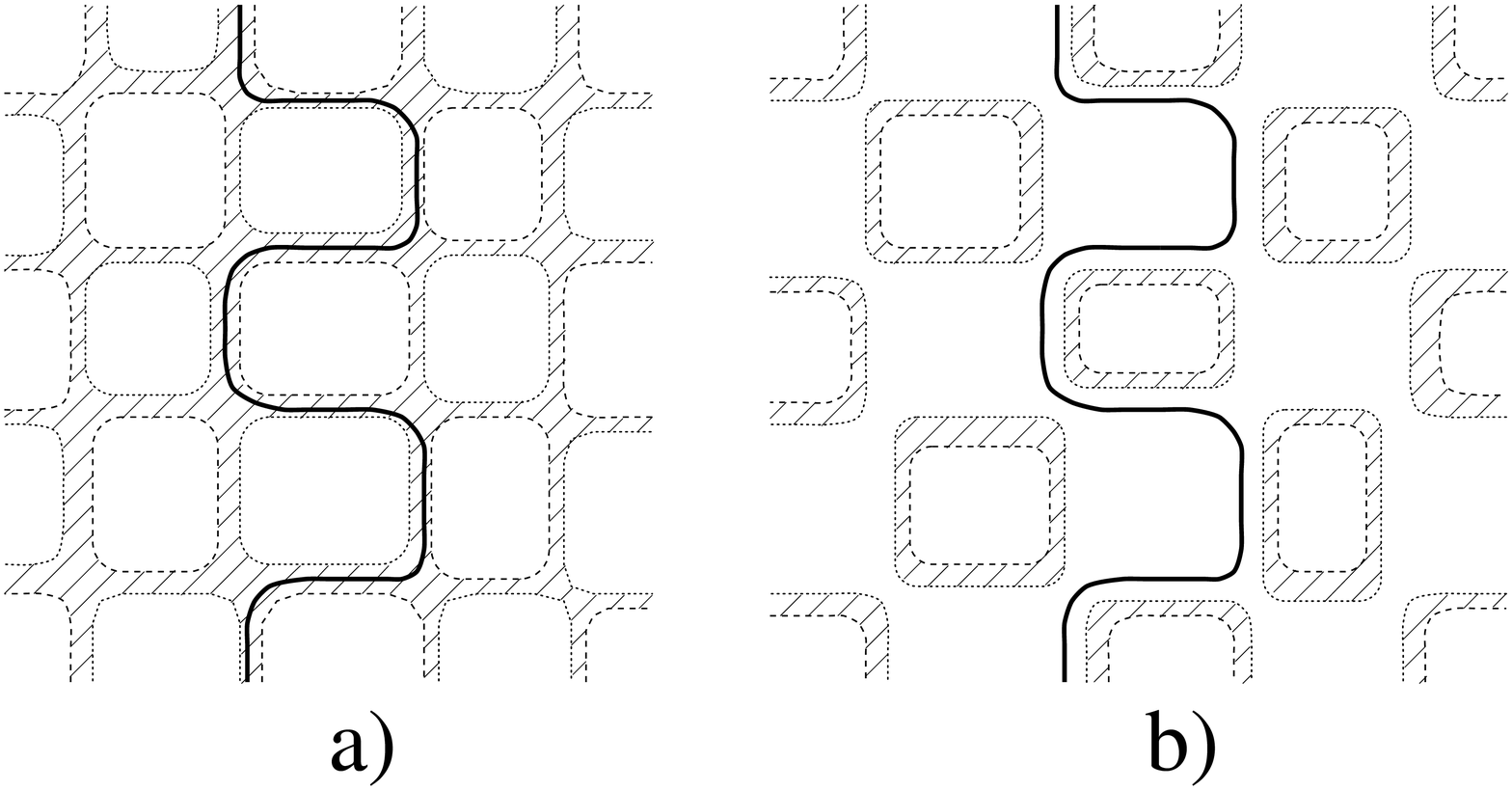}
\label{fig:web}
\baselineskip=10pt
{{\small FIG.\ 2.\quad Schematic illustration of the percolation
network in the presence of the SO-coupling at (a) $|E|<\Delta$ and (b)
$|E|>\Delta$. The bold line denotes the infinite equipotential. Dashed
is the space between trajectories with opposite spins (dotted and
dashed lines) and the same energy $E$.}}
\end{figure}
The classical treatment above is valid if the phase-breaking length
$L_\phi$, associated with inelastic scattering at finite $T$, is much
shorter than $\xi(0)$, where $\xi(0)$ is the quantum localization
length at $E=0$. We thus predict that at $L_\phi\alt\xi(0)$,
$T\ll\Delta$, and $L_{so}\alt L(\Delta)$ the dissipative conductivity
$\sigma_{xx}$ as a function of the Fermi energy $E_F$ exhibits a
box-like behavior [Fig.\ 3]: $\sigma_{xx}\sim e^2/h$ inside the band
$|E_F|<\Delta$ and is exponentially small otherwise. The point is that
for $|E_F|>\Delta$ the conductivity is only due to activation to the
nearest percolation threshold, while a {\it metallic band with
well-pronounced boundaries} exists between the percolation
thresholds. The easiest way to see that $\sigma_{xx}\sim e^2/h$ at
$|E_F|<\Delta$ is as follows. The diffusion coefficient $D$ provided
by the SO-coupling of neighboring clusters is of order $R^2
(\Delta)v_d/L(\Delta)$, where $v_d$ is the typical drift velocity. On
the other hand, the density of electron states on the loop of the
length $L(\Delta)$ is $g\sim [1/R^2(\Delta)]L(\Delta)/hv_d$. Thus, if
the SO-coupling is strong enough, i.e. $L_{so}\alt L(\Delta)$, we have
$\sigma_{xx}=e^2gD\sim e^2/h$. In the absence of the SO-coupling, the
wave functions decay on the scale of $\lambda$ from the equipotentials
and so the conductivity inside the band $|E_F|<\Delta$ would be
dominated by activation as long as $T\ll\Delta$. Let us stress that
both conditions $L_\phi\alt\xi(0)$ and $T\ll\Delta$ are necessary for
the box-like behavior of $\sigma_{xx}$. In this Letter we do not
specify whether $L_\phi$ is limited by electron-electron or
electron-phonon scattering and keep $L_\phi$ as a phenomenological
length. We wish to note, however, that in any case the conditions can
be fulfilled simultaneously at small enough Zeeman splitting as
$\xi(0)$ grows rapidly with decreasing $\Delta$ (see below).
\begin{figure}
\epsfxsize=2.65truein
\vspace{-1.9truein}
\epsffile{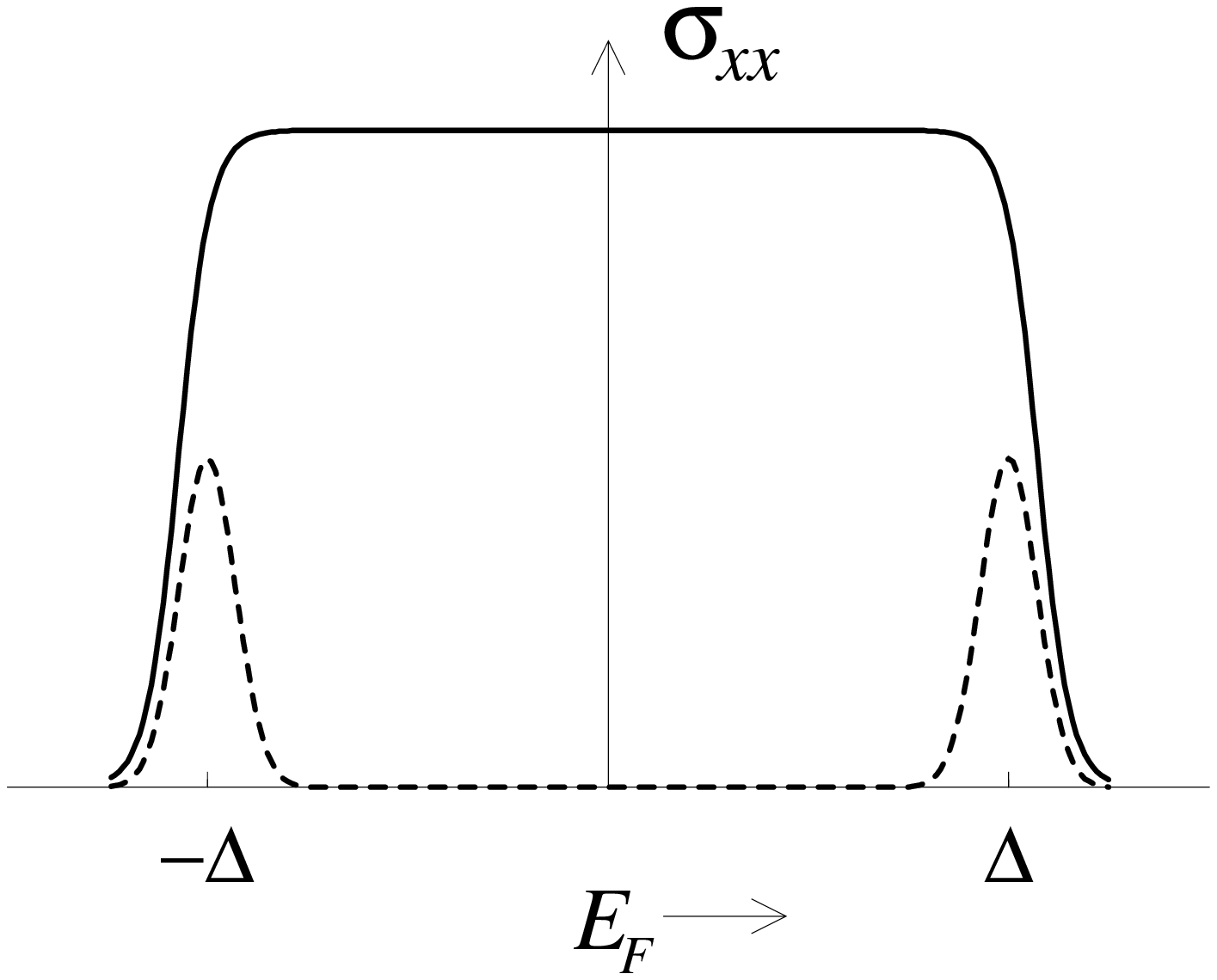}
\label{fig:box}
\baselineskip=10pt
{{\small FIG.\ 3.\quad Shown schematically is the dissipative
conductivity as a function of the Fermi level position in overlapping
disorder-broadened Zeeman levels separated by the energy $2\Delta$
with (solid line) and without (dashed line) spin-orbit coupling. Both
pictures correspond to the same temperature.}}
\end{figure} 
Now consider what happens when $L_\phi$ exceeds $\xi(0)$ with lowering
$T$. Obviously, then a drop of $\sigma_{xx}$ occurs between the
centers of the Zeeman levels and two well-pronounced
$\sigma_{xx}$-peaks appear. To estimate $\sigma_{xx}$ between the
peaks, note that owing to the SO-coupling the wave functions of the
localized states at $|E|<\Delta$ are built up from pieces of the
classical trajectories, so that the overlap integral of two states
{\it does not involve} tunneling through the saddle-points. As a
result, the wave functions overlap strongly. Accordingly,
$\sigma_{xx}$ in the middle between the peaks is due to hopping
between the states separated by $R(\Delta)$, the hopping rate being
$\sim (v_d/L(\Delta))(\xi(0)/L_\phi)$, i.e. the diffusion coefficient
is $L_\phi/\xi(0)$ times smaller than in the classical case considered
above. It follows that $\sigma_{xx}\sim (e^2/h)(\xi(0)/L_\phi)$
\cite{vrh}. Thus two peaks of $\sigma_{xx}$ as a function of $E_F$
should be strongly asymmetric: $\sigma_{xx}$ falls off rapidly at
$|E_F|>\Delta$ (activation) and slowly at $|E_F|<\Delta$ (hopping).

It is clear that the quantum localization length $\xi(0)$ resulting
from the SO-coupling is of order $R(\Delta)$ at $L_{so}\sim
L(\Delta)$. The question is whether the ratio $\xi(0)/R(\Delta)$
remains finite as $L_{so}/L(\Delta)\to 0$. The problem can be mapped
on that considered numerically in Ref.\ \cite{lee94} if the
percolation network is replaced by a square lattice with a lattice
constant $R(\Delta)$. To reconcile the model \cite{lee94} with our
picture, the amplitudes to go to the left or to the right at the nodes
of the lattice should be chosen equal to either 0 or 1 depending on
the spin-orientation. According to the numerical simulation
\cite{lee94}, the localization length in this case is about $3\times
10^2$ lattice spacings in the limit of complete spin-mixing (i.e.  it
is much larger numerically than would be anticipated on the basis of
the scaling arguments, which may be important from the experimental
point of view). Apart from the large numerical coefficient, we are now
in a position to find the energy dependence of the localization length
$\xi(E)$ at $L_{so}\alt L(\Delta)$. As $E$ approaches either of the
percolation thresholds, the characteristic radius of the critical
cluster corresponding to one of the spin projections diverges as
$R(\Delta -|E|)\sim R(\Delta)[\Delta/(\Delta -|E|)]^{4/3}$ while that
corresponding to the other remains equal to $R(\Delta)$. If we
consider only these two scales and neglect all others, the two-scale
conducting network can be modeled similarly to Ref.\
\cite{lee94}. Coupling of large cells with each other is then provided
by cells of smaller size. Clearly, the quantum localization length in
this model scales with the largest radius, $R(\Delta -|E|)$, while the
radius $R(\Delta)$ determines the distance on which the wave function
decays across the links of the large cells. In the random potential,
the topology of the percolation network is more complicated (see
above), however, the main issue remains unchanged: provided
$L_{so}\alt L(\Delta)$, the quantum localization length, which
describes {\it gradual} (as compared to the case with no SO-coupling)
fall-off of the wave function, scales as the classical percolation
radius \cite{tun}.

In conclusion, we demonstrated that, for electrons in a smooth random
potential in the QHE regime, the SO-coupling of two disorder-broadened
Zeeman levels can result in a strong suppression of the
localization. The effect shows up in a well-defined range of energies
lying between the centers of the Zeeman levels. The picture suggested
does not include effects of electron-electron interaction, such as
screening of the random potential or exchange interactions. This issue
warrants further study.


We thank J.\ Hajdu for interesting discussions. M.\ R.\ is grateful to
the Theoretical Physics Institute at the University of Cologne for
hospitality. The work was supported by the Humboldt Foundation.


\begin{references}
\bibitem{tunwidth} To estimate the width of the band, we note that, at
a given $E$, the saddle-points that separate critical
percolation-clusters have heights $\sim |E|$. According to H.\ A.\
Fertig and B.\ I.\ Halperin, \pr {\bf 36}, 7969 (1987), the tunneling
transmittancy of a saddle-point with a height $E$ and curvatures along
the principal directions $V''_x$ and $V''_y$ is $[1+\exp (2\pi
E/V''\lambda^2)]^{-1}$, where $V''=(V''_xV''_y)^{1/2}$. In the random
potential, the characteristic value of $V''$ is $\Gamma/d^2$. So the
classical picture breaks within the band $|E|\alt\Gamma (\lambda/d)^2$
where the tunneling exponent for the critical saddle-points is of
order unity.
\bibitem{zs} The SO-coupling leads to the appearance of the effective
$g$-factor which is not zero even if the ``bare" $g$-factor is
neglected. We mean by $2\Delta$ the ``net" energy distance between the
centers of the disorder-broadened Zeeman levels, which is determined
by the SO-interaction as well.
\bibitem{loop} It is worth noting that other saddle-points with
heights $\sim |\Delta|$ which they meet on their way to the critical
one do not change the estimate.  Though their number, $\sim
L(\Delta)\delta/d^2\sim R(\Delta)/d$, is large, typically they just
connect up loops of size $d$. (The probability density of the loop
perimeter falls off as a power-law function whose exponent, 11/7, can
be found from the condition that there is only one loop of the maximum
length $L(\Delta)$.) Since the average distance between the
saddle-points along the trajectories, $\sim d^2/\delta$, is much
larger than $d$, only a small fraction of them form multiply connected
figures.
\bibitem{vrh} This law is violated with decreasing $T$ only when $T$
becomes so low that the typical energy spacing between neighboring
states exceeds $T$, then the mechanism of transport changes to the
variable-range-hopping.
\bibitem{tun} The law $\xi(E)\sim R(\Delta -|E|)$ is valid between the
percolation thresholds as long as $\Delta -|E|\agt\Delta_t$. Within
the energy bands of the width $\Delta_t$ around the thresholds,
tunneling between the critical clusters of size $R(\Delta_t)$ comes
into play, so that the presence of larger clusters can be ignored and
the characteristic size of the conducting-network cell, $R(\Delta_t)$,
is independent of $E$.
\end{references}
\end{document}